\documentclass[11pt,british,notitlepage]{revtex4-1}
\usepackage[LGR,T1]{fontenc}
\usepackage[latin9]{inputenc}
\usepackage[a4paper]{geometry}
\usepackage{color}
\usepackage{textcomp}
\usepackage{amsmath}
\usepackage{amssymb}
\usepackage{graphicx}
\usepackage{setspace}
\usepackage{gensymb}

\makeatletter

\DeclareRobustCommand{\greektext}{%
  \fontencoding{LGR}\selectfont\def\encodingdefault{LGR}}
\DeclareRobustCommand{\textgreek}[1]{\leavevmode{\greektext #1}}
\ProvideTextCommand{\~}{LGR}[1]{\char126#1}

\usepackage{lineno}

\makeatother

\usepackage{babel}
\begin{document}
%\linenumbers
\title{Controllable Josephson junction for photon Bose-Einstein condensates}
\author{Mario Vretenar, Ben Kassenberg, Shivan Bissesar, Chris Toebes, and
Jan Klaers\vspace{1.5mm}
}
\address{Complex Photonic Systems (COPS), MESA$^{+}$ Institute for Nanotechnology,
University of Twente, PO Box 217, 7500 AE Enschede, Netherlands}
\begin{abstract}
\noindent Josephson junctions are the basis for the most sensitive
magnetic flux detectors, the definition of the unit volt by the Josephson
voltage standard, and superconducting digital and quantum computing.
They result from the coupling of two coherent quantum states, as they
occur in superconductors, superfluids, atomic Bose-Einstein condensates,
and exciton-polariton condensates. In their ground state, Josephson
junctions are characterised by an intrinsic phase jump. Controlling
this phase jump is fundamental for applications in computing. Here,
we experimentally demonstrate controllable phase relations between
photon Bose-Einstein condensates resulting from particle exchange
in a thermo-optically tunable potential landscape. Our experiment
realises an optical analogue of a controllable 0,$\pi$-Josephson
junction. \textcolor{black}{By connecting several junctions, we can
study a reconfigurable 4-condensate system demonstrating the potential
of our approach for analog spin glass simulation. }More generally,
the combination of static and dynamic nanostructuring techniques introduced
in our work offers a powerful platform for the implementation of adaptive
optical systems for paraxial light in and outside of thermal equilibrium.
\end{abstract}
\maketitle

\section{Introduction}

Finding the energetic ground state of a magnet with disordered couplings
is a complicated combinatorial problem. This so called spin glass
problem has \textcolor{black}{no analytic solution and even numerical
techniques are found to be inefficient. It is known that many important
optimisation problems in machine learning, logistics, computer chip
design and DNA sequencing can be mathematically mapped to an equivalent
spin glass problem \citep{Luc14}. The latter results from the proven
NP hardness of the problem \citep{Bar82,Cub16}. A method for solving
the spin glass problem in the Ising, XY or Heisenberg model can, thus,
serve as a blueprint for approaching a large class of mathematical
optimisations. This motivates research on analog spin glass simulators
as a new class of computational devices specifically designed for
simulating spin systems. For the Ising model, a first generation of analog spin glass simulators
has been realized using networks of superconducting qubits \citep{Joh11,Boi14}
and optical parametrical oscillators \citep{Mar14,McM16,Ina16}. Whether these simulators already offer a computational advantage over the more conventional von Neumann computer architecture is currently a matter of controversy. 
For the XY model, the proposed physical platforms are based on superconducting
qubits \citep{Kin18}, lasers \citep{Nix13}, atomic Bose-Einstein condensates (BEC) \citep{Str13}
and polariton condensates \citep{Oha16,Ber17,Kal18,Oha18,Kal19,Aly20}.
The basic idea in these approaches is to associate the XY spins $\varphi_{i}\in[0,2\pi)$
to the phases of coherent states. A prototypical spin glass simulator
consists of a lattice of simulated spins, which are coupled to each
other in a controllable manner. In the case of coupled BECs, the couplings
can be considered as Josephson junctions \citep{Jos62,Cat01,Lai07,Lag10,Abb13,Gin16,Adi17}. 
The ability to accurately adjust the coupling constants between the
spins is fundamental for defining the computational problem to be
solved. In earlier work with polariton condensates, tunable couplings
were realised by controlling the geometrical distance between the
condensates or by exploiting polariton-reservoir interactions \citep{Oha16,Ber17,Oha18,Aly20}.
The former approach is mainly limited to systems with homogenous couplings
across the lattice. Polariton-reservoir and polariton-polariton interactions
generally result in a gain \citep{Oha16,Oha18} and time dependence
\citep{Oha16} of the condensate couplings. This is undesirable for
spin glass simulation as this leads to a continuous redefinition of
the computational problem when the system is amplified from the quantum
to the classical regime by increasing the optical gain. In this work,
we experimentally realise a controllable $0,\pi$-Josephson junction
for photon Bose-Einstein condensates exploiting the anomalously large
thermo-optical coefficient of an optical medium close to a phase transition.
By connecting several junctions, we can study effective 4-spin systems
demonstrating the potential of our approach for analog spin glass
simulation.}

\section{Experimental system}

\textcolor{black}{}
\begin{figure}
\begin{centering}
\textcolor{black}{\includegraphics[width=17cm]{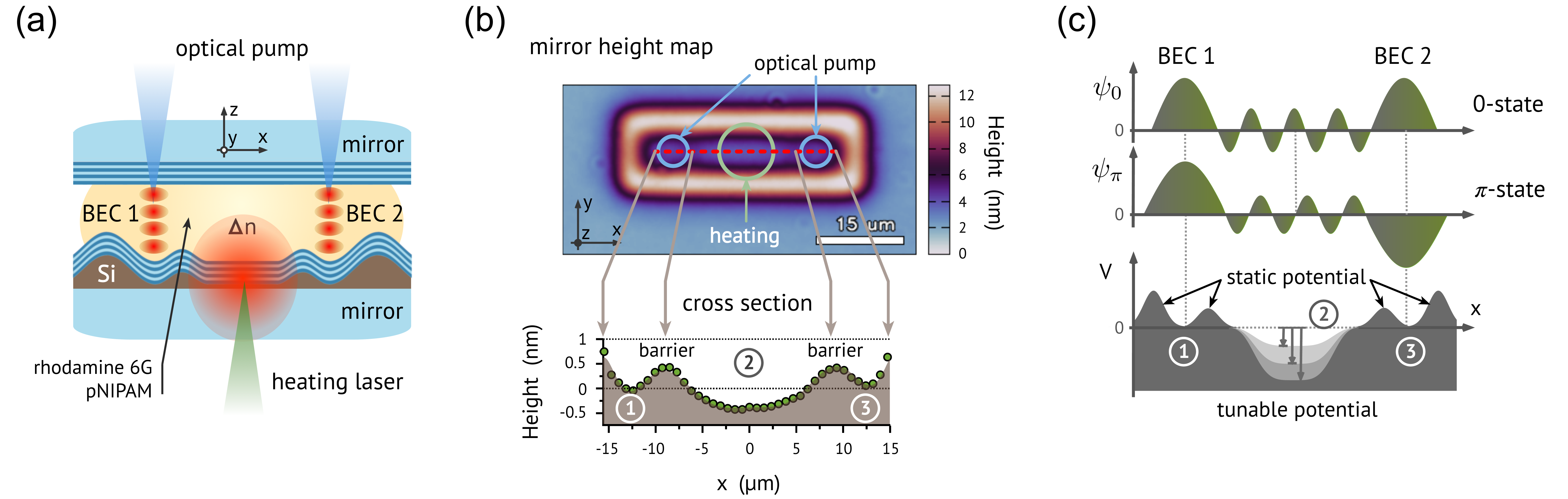}}
\par\end{centering}
\centering{}\textcolor{black}{}%
\begin{minipage}[t]{0.95\columnwidth}%
\begin{singlespace}
\textcolor{black}{\caption{Controllable Josephson junction for photon Bose-Einstein condensates.
(a) The experimental setup is based on a high-finesse dye microcavity,
in which optical photons propagate paraxial to the optical axis ($z$-axis)
and are repeatedly absorbed and reemitted by dye molecules. The latter
leads to a thermalisation and condensation of the photon gas at room
temperature. The potential landscape for the two-dimensional photon
gas is created by a combination of two experimental techniques: a
static nanostructuring of one of the mirror surfaces and an in-situ
variation of the index of refraction of the optical medium using a
thermo-responsive polymer (pNIPAM) and a heating laser. (b) Static
confinement potential. A rectangular surface structure is created
on one of the cavity mirrors, which restricts the flow of light to
the confined area. This area is furthermore divided into three parts
by introducing two shallower barriers, see the cross section in the
lower graph. The height map of the mirror was determined by Mirau
interferometry. (c) The Josephson junction in our experiment consists
of two photon Bose-Einstein condensates that are created at the two
ends of the confining potential. By tunnelling into the central region
of the junction, a particle exchange is created that leads to in-phase
or anti-phase relations between the condensates (upper two graphs).
The acquired phase delay is controlled by the thermo-optically induced
potential (lower graph).}
}
\end{singlespace}
\end{minipage}
\end{figure}
\textcolor{black}{Our experimental setup is based on a high-finesse
dye microcavity, see Fig. 1a, in which optical photons }propagate
paraxial to the optical axis ($z$-axis) \textcolor{black}{and are
repeatedly absorbed and reemitted by dye molecules. The dye molecules
obey the Kennard-Stepanov law connecting the (broadband) Einstein
coefficients of absorption $B_{12}(\omega)$ and emission $B_{21}(\omega)$
by a Boltzmann factor: $B_{12}(\omega)/B_{21}(\omega)=\exp[\hbar(\omega\hspace{-0.5mm}-\hspace{-0.5mm}\omega_{\text{zpl}})/kT]$.
Here, $\omega_{\text{zpl}}$ is the zero-phonon line of the dye and
$T$ is the temperature. Multiple absorption-emission cycles establish
a thermal contact between the photon gas and its environment, which,
in effect, leads to a thermalisation and condensation of the photon
gas at room temperature \citep{Kla10b,Kla10,Kir13,Dun17,Wal18,Gre18,Gla20}.
For sufficiently small mirror spacings, the photon gas effectively
becomes two-dimensional and follows a modified energy-momentum relation
given by
\begin{equation}
E\simeq\frac{mc^{2}}{n_{0}^{2}}+\frac{(\hbar k_{r})^{2}}{2m}-\frac{mc^{2}}{n_{0}^{2}}\left(\frac{\Delta d}{D_{0}}+\frac{\Delta n}{n_{0}}\right)\;\text{,}\label{eq:energy_momentum}
\end{equation}
where $k_{r}$ represents the transverse wavenumber and $m$ denotes
an effective photon mass (see Appendix A). The second and third term
correspond to the kinetic and potential energy of the photons. The
potential energy is non-vanishing, if the distance between the mirrors
$D(x,y)=D_{0}+\Delta d(x,y)$ or the refractive index $n(x,y)=n_{0}+\Delta n(x,y)$
is modified across the transverse plane of the resonator. In our derivation,
we assume $\Delta d\ll D_{0}$ and $\Delta n\ll n_{0}$.}

\textcolor{black}{Controlling the potential landscape within the microresonator
through static and dynamic nanostructuring techniques is fundamental
to this experiment. Various experimental techniques for shaping the
potential energy profiles in microcavity systems have been developed,
for example methods based on strain \citep{Bal07}, surface waves
\citep{Lim06}, electrostatic \citep{Fra11} and exciton reservoir
confinement \citep{Wer10}, and deep etching \citep{Mor01}. The Josephson
junction in our experiment consists of two photon Bose-Einstein condensates that exchange particles in a tunable potential landscape created by a combination of two experimental techniques: nanostructuring of the mirror surface and in-situ variation of the index of refraction
of the optical medium. For the static nanostructuring, we use a novel direct
laser writing technique that allows us to selectively lift the surface
of the mirror by several hundred nanometers with sub-nanometer control
\citep{Kur20}. This translates into repulsive potentials for the
photon gas in a microresonator setting, see eq. (\ref{eq:energy_momentum}).
In a first step, we create a barrier around a rectangular region of
$30\,\mu\text{m}$ length and $5\,\mu\text{m}$ width, see Fig. 1b,
which restricts the flow of light to the confined area. In a second
step, we divide the confined area into three regions by introducing
two shallower barriers (the three regions are denoted by 1, 2, and
3 in Fig. 1b,c). The microresonator setup is completed by a second
(planar) mirror and an optical medium, which is a water-based solution
of rhodamine 6G and the thermo-responsive polymer pNIPAM. By optically
pumping the dye molecules with a $5\,\text{ns}$ laser pulse at $\lambda=480\,\text{nm}$,
we create photon Bose-Einstein condensates at the two ends of the
structure. The finite potential barriers allow photons to tunnel into
the central region of the junction establishing a particle exchange
between the two condensates. The thermo-responsive polymer pNIPAM
is used to control the potential landscape in the central region of
the junction. For that, we use a second laser that emits $2\,\text{ms}$
long laser pulses of variable energy. These pulses are irradiated
onto the sample $50\,\text{ms}$ before the optical pumping initiates
the condensation process. The pulses are absorbed in an amorphous
silicon layer located below the dielectric stack of one of our mirrors
('Si' in Fig. 1a). The absorbed energy increases the local temperature
of the optical medium by a few Kelvin such that it reaches the lower
critical solution temperature (LCST) of pNIPAM in water (32 \degree C). This
leads to a significant change of the index of refraction which, with
eq. (\ref{eq:energy_momentum}), translates into a tunable potential
for the photons in the microcavity \citep{Dun17}. The purpose of
this tunable potential is to control the phase delay that the photons
acquire while traveling through the central region of the junction, see Fig. 1c.}

\textcolor{black}{The coupled photon BEC system follows equations
of motion that closely resemble the Josephson equations (see Appendix
B). Deviations from the textbook Josephson scenario arise from the
interaction with the environment. In the photon BEC system, the states
of the condensates are not given entities but are subject to energy
and particle exchange with the environment. In particular, the evolution
of the total photon number $n=n_{1}+n_{2}$ is equal to
\begin{equation}
\dot{n}=\lambda\,n-J_{12}\sqrt{n_{1}n_{2}}\,\cos(\theta_{1}\hspace{-0.5mm}-\hspace{-0.5mm}\theta_{2})\;\text{,}\label{eq:gain}
\end{equation}
where $n_{1,2}$ and $\theta_{1,2}$ are the photon numbers and phases
of the two condensates, $\lambda$ denotes a gain parameter and $J_{12}$
is the coupling between the two condensates (see Appendix B). Upon
optically pumping the system, the condensation process is triggered
and a competition among the possible system states is initiated. In
this competition, the state that maximises the gain $\dot{n}$ is
the one that acquires the dominant statistical weight. Based on eq.
(\ref{eq:gain}), we expect that the system predominantly realises
states with equal condensate population (as this maximises the square
root function for a given $n$) and phase differences of $\theta_{1}-\theta_{2}=0$
($0$-state) or $\theta_{1}-\theta_{2}=\pi$ ($\pi$-state) depending
on the sign of $J_{12}$. The latter is a function of the induced
potential and other system parameters (see Appendix B). These two
states maximise the gain, are fixed points of the Josephson equations
and, thus, define the intrinsic phase jump across the junction.}

\section{Results}

\subsection{Single junction}

\textcolor{black}{}
\begin{figure}
\begin{centering}
\textcolor{black}{\includegraphics[width=17cm]{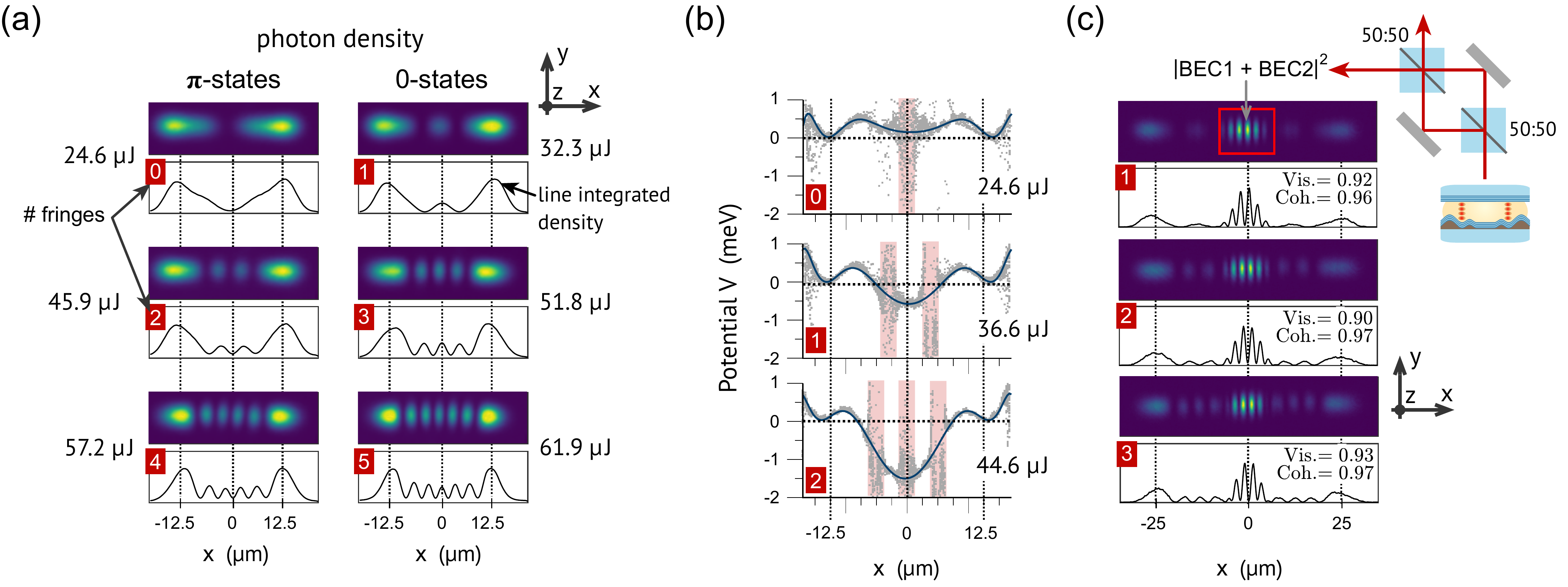}}
\par\end{centering}
\centering{}\textcolor{black}{}%
\begin{minipage}[t]{0.95\columnwidth}%
\begin{singlespace}
\textcolor{black}{\caption{Controllable phase relations. (a) Photon density in the microresonator
plane for six different heating energies as determined by a camera
capturing the light transmitted by one of the cavity mirrors. All
density profiles can be assigned to symmetric wavefunctions ($0$-states)
or antisymmetric wavefunctions ($\pi$-states). The numbers in the
red boxes indicate the number of observed intensity maxima between
the condensates. (b) Potential landscape in the microresonator. Starting
from the measured photon density $\rho=|\psi(x,y)|^{2}$, we reconstruct
the wavefunction of the photons $\psi(x)=\epsilon(x)\sqrt{\rho(x)}$,
in which $\epsilon(x)$ switches between $+1$ and $-1$ at every
node in the wavefunction. The potential $V$ follows via $V-\text{const}=-E_{\text{kin}}=(\hbar^{2}/2m)(d^{2}\psi/dx^{2})\psi^{-1}$
(points). Our method breaks down at the nodes of the wavefunction.
These regions (shaded areas) are excluded from the polynomial fit
(line). The reconstructed potentials reveal both the static potential
due to the nanostructured mirror and the dynamically induced potential
due to the thermo-responsive polymer. (c) Interferometric images of
the coupled condensate system. The transmitted light is guided through
a Mach-Zehnder interferometer to test for the coherence of the two
condensates. Data for the interferometric visibility $v=(I_{\text{\text{max}}}-I_{\text{\text{min}}})/(I_{\text{\text{max}}}+I_{\text{\text{min}}})$
is given in the inset. The estimated degree of first-order coherence,
which is generally found to be close to unity, takes into account
the (small) population imbalances between the condensates.}
}
\end{singlespace}
\end{minipage}
\end{figure}
\textcolor{black}{}
\begin{figure}
\begin{centering}
\textcolor{black}{\includegraphics[width=8cm]{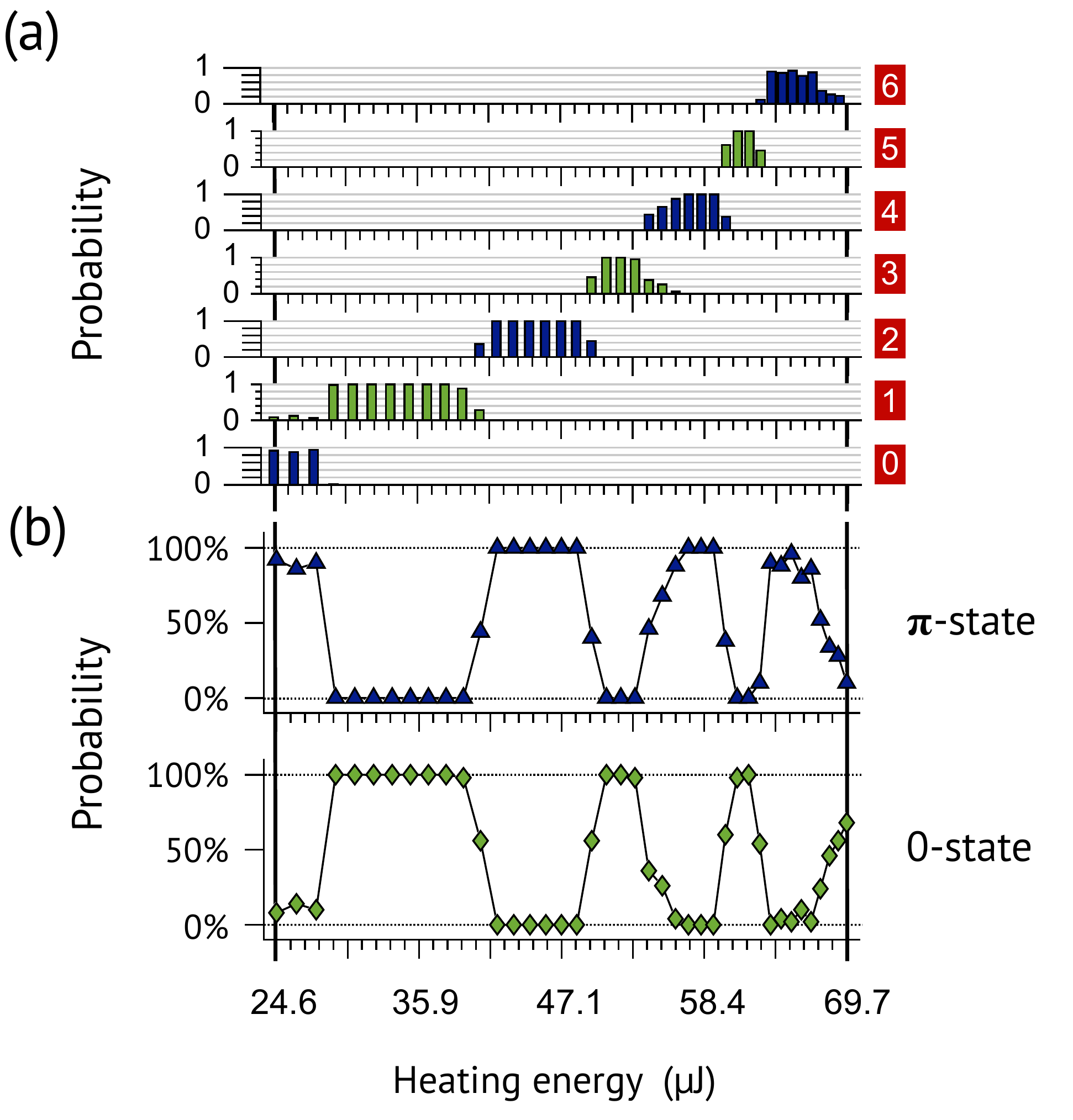}}
\par\end{centering}
\begin{centering}
\textcolor{black}{\vspace{-4mm}
}%
\begin{minipage}[t]{0.95\columnwidth}%
\begin{singlespace}
\textcolor{black}{\caption{State-resolved statistics. (a) State-resolved statistical information
on the operation of the junction. For every heating energy, 50 experimental
runs with identical control parameters are measured and analysed.
The data indicates that the formation of a particular junction state,
as defined by the number of observed intensity maxima between the
condensates (red boxes), is deterministic for a significant fraction
of the investigated heating energies. \textbf{b}, Probability of finding
the junction in a $0$- or $\pi-$state as a function of the heating
energy.}
}
\end{singlespace}
\end{minipage}
\par\end{centering}
\begin{onehalfspace}
\textcolor{black}{\vspace{2mm}
}
\end{onehalfspace}
\end{figure}
\textcolor{black}{Figure 2a shows the photon density $\rho=|\psi(x,y)|^{2}$
inside the microresonator as determined by a camera capturing the
light transmitted by one of the cavity mirrors. We observe various
stripe patterns indicating the presence of photon exchange and the
formation of standing waves between the condensates. All density profiles
can be clearly assigned to symmetric wavefunctions ($0$-states) or
antisymmetric wavefunctions ($\pi$-states), see Fig. 1c. With increasing
potential depth, symmetric and antisymmetric states alternate and
the number of nodes in the wavefunction increases. The measured photon
density $\rho$ allows us to reconstruct the potential landscape in
the microresonator. Using $V-\text{const}=-E_{\text{kin}}=(\hbar^{2}/2m)(d^{2}\psi/dx^{2})\psi^{-1}$
with $\psi(x)=\epsilon(x)\sqrt{\rho(x)}$ and a sign function $\epsilon(x)$
switching between $+1$ and $-1$ at every node, we obtain the spatial
variation of the potential in the $x$-direction of the microcavity
plane, see Fig. 2b. The reconstructed potentials clearly reveal both
the static potential due to the nanostructured mirror and the dynamically
induced potential due to the thermo-responsive polymer. Interferometric
images of the coupled condensate system are shown Fig. 2c. For this,
we guide the transmitted light through a Mach-Zehnder interferometer,
see the schematic representation in Fig. 2c, and superimpose the images
of the two condensates. The high contrast of the observed stripe pattern
indicates first-order coherences close to unity. The fact that this
level of coherence is reached when integrating over one gaussian pump
pulse proves that the coupling in our junction is neither dependent
on optical gain nor influenced by self-interactions. The latter is
an essential requirement for using the junction in optical spin glass
simulation. State-resolved statistical information on the operation
of the junction are provided in Fig. 3a. Here, we measure and analyse
sequences of 50 experimental runs with identical control parameters.
The data indicates that the formation of a particular phase relation
between the condensates is deterministic for a significant fraction
of the investigated heating energies. Moreover, the transition between
the states is quite sharp, see Fig. 3b.}

\subsection{Multiple junctions}

\textcolor{black}{}
\begin{figure}
\begin{centering}
\textcolor{black}{\includegraphics[width=17cm]{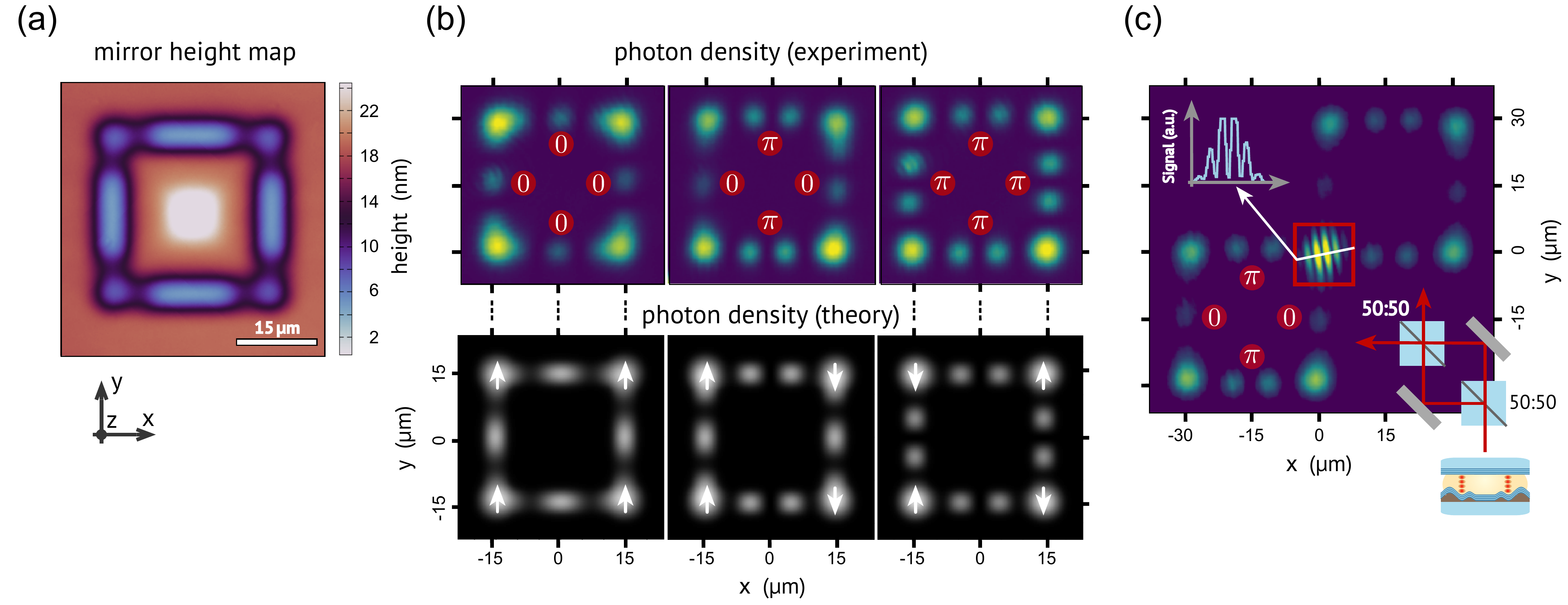}\vspace{-4mm}
}%
\begin{minipage}[t]{0.95\columnwidth}%
\begin{singlespace}
\textcolor{black}{\caption{Higher-order optical circuit composed of four Josephson junctions.
(a) Height map of the used cavity mirror. Optical pumping with nanosecond
pulses is carried out at the four corners of the structure. (b) Photon
density for three different coupling configurations as denoted by
the labels '0' and '$\pi$'. By varying the spatial heat profile in
the microcavity, the phase relations of the four condensates can be
switched between in-phase and anti-phase. The upper panel shows experimental
results, while the bottom panel contains theoretical results derived
from the numerical solution of a driven-dissipative Schr\"odinger equation
(see Appendix B). The white arrows correspond to the phases of the
photonic wavefunction at the position of the condensates and can be
interpreted as the angular orientation of four XY spins. (c) Interferometric
imaging of the four coupled condensates using a Mach-Zehnder interferometer.
The top right condensate in one interferometer path is superimposed
on the bottom left condensate in the other interferometer path. The
observed interference fringes indicate coherence close to unity (see
cross section).}
}
\end{singlespace}
\end{minipage}
\par\end{centering}
\begin{onehalfspace}
\textcolor{black}{\vspace{0mm}
}
\end{onehalfspace}
\end{figure}
\textcolor{black}{When several junctions are connected together, the
gain function eq. (\ref{eq:gain}) can be generalised to $\dot{n}=\sum_{i}\lambda\,n_{i}-H_{XY}$
with $H_{XY}=-\sum_{i>j}J_{ij}^{XY}\,\cos(\theta_{i}\hspace{-0.5mm}-\hspace{-0.5mm}\theta_{j})$
and $J_{ij}^{XY}=-J_{ij}\sqrt{n_{i}n_{j}}$. Maximising the gain thus
corresponds to minimising the energy of a simulated XY system with
couplings $J_{ij}^{XY}$, which demonstrates the connection with the
spin glass problem (see also Appendix B). Figure 4a shows a mirror surface profile that
is designed to implement four mutually coupled Josephson junctions with rectangular
symmetry. When the four corners of the structure are pumped optically,
condensates form and establish phase relations with one another. By
triggering the polymer phase transition, we can switch between positive
(in-phase) and negative (anti-phase) couplings in this structure.
Three different states of the condensate lattice, which result from
three different heat patterns applied to the thermo-responsive optical
medium, are shown in Fig 4b. In all cases, we have verified that
the four condensates are coherent to each other, as exemplarily shown
in Fig. 4c. The observed states can be interpreted as the solutions
to three different ground state problems defined in a 4-spin XY model.}

\section{Discussion}

\begin{figure}
\begin{centering}
\includegraphics[width=17cm]{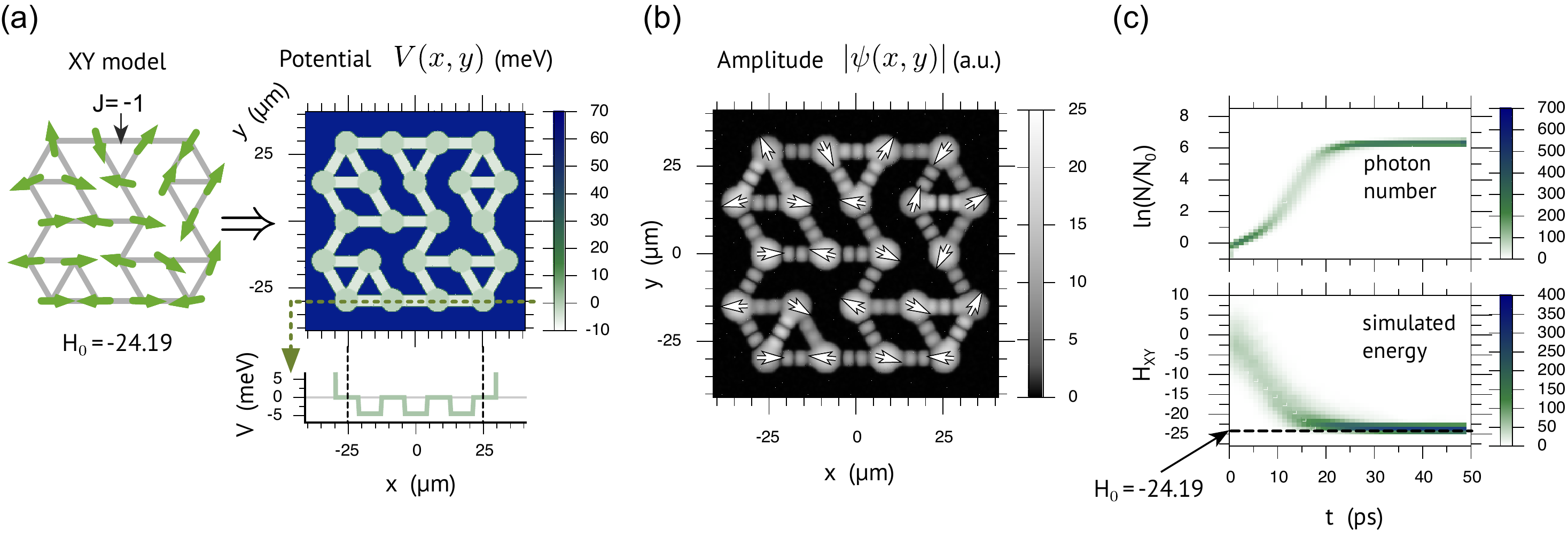}
\par\end{centering}
\begin{centering}
\vspace{-4mm}
\begin{minipage}[t]{0.95\columnwidth}%
\begin{singlespace}
\caption{Proposed scheme for optical spin glass simulation with photonic Josephson
junctions. (a) Mapping between XY spin glasses and systems of coupled
Bose-Einstein condensates. We propose to use photonic Josephson junctions
to solve optimisation problems, namely the ground state problem in
XY spin glasses and mathematical equivalent problems. The ground state
problem corresponds to finding the minimum of $H_{XY}=-\sum_{i>j}J_{ij}\cos(\phi_{i}-\phi_{j})$
with optimisation variables $\phi_{i}\in[0,2\pi)$. A particular instance
of such a problem is defined on a (random) 3-regular planar graph
with antiferromagnetic couplings $J_{ij}=-1$. The problem is mapped
onto the potential landscape in the microcavity system by introducing
a triangular lattice potential, in which the lattice sites are connected
by attractive step potentials. The depth of the steps is chosen to
favour the formation of $\pi$-states. (b) Amplitude of the simulated
light field $|\psi(x,y)|$. We have performed numerical simulations
to investigate the condensation process in the given potential landscape.
A two-dimensional stochastic Schr\"odinger equation with gain and
frequency-dependent loss terms is numerically integrated using the
Runge-Kutta method (see Appendix B). At the start of the simulation,
the optical gain on the triangular lattice is set close to the condensation
threshold and photons start to populate the cavity. The phases of
the light field, indicated by the arrows, are adjusted to maximise
the gain. (c) Total photon number and simulated energy as a function
of time. The data comprises 300 numerically obtained stochastic time
evolutions of the system. The relative frequency of the total photon
number $N$ and the simulated energy $H_{XY}$, derived from the phases
of the light field, are colour-coded. The photon number is normalised
to the noise level $N_{0}$ in the stochastic Schr\"odinger equation
(at zero gain). The coupled condensate system is capable of finding
good approximative solutions of the given ground state problem within
a period as short as $20\,\text{ps}$.}
\end{singlespace}
\end{minipage}
\par\end{centering}
\begin{onehalfspace}
\vspace{0mm}
\end{onehalfspace}
\end{figure}
We expect that analog spin glass simulation will be the first major
application of photonic Josephson junctions. To illustrate the potential
of this approach, we propose and numerically analyse a scheme for
solving the ground state problem in XY spin glasses using networks
of coupled photon Bose-Einstein condensates. A particular instance of such a problem
is defined in Fig. 5a, which shows a classical XY model with Hamiltonian
$H_{XY}=-\sum_{i>j}J_{ij}\cos(\phi_{i}-\phi_{j})$ on a (random) 3-regular
planar graph with antiferromagnetic couplings $J_{ij}=-1$. As discussed
before, the basic idea is to associate the XY spins $\phi_{i}\in[0,2\pi)$
to the phases $\theta_{i}$ of photon Bose-Einstein condensates. For
the given example, we assume a triangular lattice potential in the
microcavity. The lattice sites are connected to each other by attractive
step potentials with variable depth. For mapping the XY coupling constants
$J_{ij}=-1$ to the photon BEC system, we set the potential depth
to a value that favours the formation of $\pi$-states. This results
in the potential landscape shown in Fig. 5a. We have performed numerical
simulations to investigate the dynamics of the condensation process
in the given potential landscape. For this, we numerically integrate
a two-dimensional stochastic Schr\"odinger equation with gain and
frequency-dependent loss terms in an experimentally relevant parameter
regime (see Appendix B). At $t=0$, the optical gain on the triangular
lattice is set close to the condensation threshold and photons start
to populate the cavity. This starts a competition among the various
system states, in which the phases of the light field, indicated by
the arrows in Fig. 5b, are adjusted to maximise the gain function.
As discussed before, the latter is a straightforward generalisation of eq. (\ref{eq:gain})
and can be shown to coincide with the XY Hamiltonian $H_{XY}$ (see
Appendix B). Figure 5c shows the total photon number and the simulated
energy, derived from the phases of the light field, as a function
of time. The data comprises 300 numerically obtained stochastic time
evolutions of the system. The relative frequency of photon number
and simulated energy are colour-coded. From this figure, we conclude
that the coupled condensate system is capable of finding good approximative
solutions of the given ground state problem within several ten picoseconds.
This is faster than a single tick of the clock in a conventional CPU,
which gives an indication of the potential of optical spin glass simulation.
That said, we would like to emphasise that many aspects determining
the performance of the proposed simulator will need to be investigated
more closely in the future and several technological challenges have yet to be overcome before such investigations can begin.

\section{Conclusion}

In conclusion, our work introduces and investigates
Josephson junctions for photon Bose-Einstein condensates controlled by thermo-optical interactions.
We expect that photonic Josephson junctions can play a major role
for novel computational schemes, for example, in analog spin glass
simulation and oscillatory neural networks. Various approaches to realizing such simulators are currently being investigated. At this time it is not clear whether the first generation of spin glass simulators provides a computational advantage over the more conventional von Neumann computer architecture. In the different approaches there are also different challenges that have to be overcome, for example, the replacement of digital electronic components in the computation process, miniaturization, non-local couplings, and scaling. At the moment, we do not see a system that has overcome all of these challenges. For the photon BEC system, scaling to larger system sizes is the most challenging aspect. In general, phase-coherent coupling of several hundreds or thousands of photon condensates is technically feasible and has already been achieved for photon BECs and related systems. However, controllable couplings with single bond resolution have not yet been realized in such systems and remain a challenge. The research on Bose-Einstein condensation of photons has brought forth experimental techniques that allow an unprecedented level of control of microresonators  \citep{Kur20,Dun17}. Based on the results presented in this work, we believe that these techniques are suitable for overcoming this challenge. Beyond the application in spin glass simulation, the combination of static nanostructuring with a thermo-responsive optical medium introduced in this work moreover offers a powerful platform for the implementation of adaptive optical circuits, lattice systems, and photonic crystals for paraxial light.

\section*{Acknowledgment}

\textcolor{black}{We thank Klaas-Jan Gorter for his contributions
in the early phase of this project and the Weitz group at the University
of Bonn for providing critical equipment. Useful discussions with
Natalia Berloff, Pavlos Lagoudakis, Michiel Wouters, Pepijn Pinkse,
and Willem Vos are acknowledged.}

\nolinenumbers

\section*{APPENDIX A: EXPERIMENTAL METHODS\vspace{-3mm}
}

\subsection*{1. Microcavity set-up\vspace{-3mm}
}

Our experiment is based on a high finesse microcavity as shown in
Fig. 1a of our Letter. In this system, photons behave as two-dimensional
massive particles subject to a controllable potential energy landscape.
In general, the photon energy in the cavity is given by $E=(\hbar c/n)\sqrt{k_{z}^{2}+k_{r}^{2}}$,
in which the longitudinal (along the optical axis) and transverse
wavenumbers are denoted by $k_{z}$ and $k_{r}$. The boundary conditions
induced by the mirrors require $k_{z}=q\pi/D(x,y)$, in which $q=\text{const}$
is the longitudinal mode number and $D(x,y)$ denotes the mirror separation.
We allow small variations in the index of refraction $n(x,y)=n_{0}+\Delta n(x,y)$
and in the mirror separation $D(x,y)=D_{0}+\Delta d(x,y)$ across
the transverse plane of the resonator. Assuming $k_{r}\ll k_{z}$,
$\Delta n\ll n_{0}$ and $\Delta d\ll D_{0}$, we can approximate
the photon energy by 
\begin{equation}
E\simeq\frac{mc^{2}}{n_{0}^{2}}+\frac{(\hbar k_{r})^{2}}{2m}-\frac{mc^{2}}{n_{0}^{2}}\left(\frac{\Delta d}{D_{0}}+\frac{\Delta n}{n_{0}}\right)\;\text{,}\label{eq:energy_photon}
\end{equation}
in which only leading order contributions were retained. Furthermore,
we have defined an effective photon mass $m=\pi\hbar n_{0}q/cD_{0}$.
The first term corresponds to the rest energy of the (two-dimensional)
photon. The second term is the kinetic energy and the third term corresponds
to the potential energy of the photon. The latter is non-vanishing,
if the index of refraction or the mirror distance varies across the
transverse plane of the resonator. In our experiment, the index of
refraction is controlled by means of a thermo-responsive polymer \citep{Dun17}
and variations of the mirror distance are achieved by nanostructuring
one of the mirrors with a direct laser writing technique \citep{Kur20}.
The height profiles of our mirrors are determined via Mirau interferometry.
For the latter, we use a commercially available interferometric microscope
objective (20X Nikon CF IC Epi Plan DI).

In our experiment, the rest energy of the photon is $mc^{2}/n_{0}^{2}\simeq2.1\,\text{eV}$
(yellow spectral regime). The mirror separation $D_{0}$ can be determined
using eq. (\ref{eq:energy_photon}), which relates height differences
to differences in potential energy: $\Delta V=-(mc^{2}/n_{0}^{2})\,\Delta d/D_{0}$.
Based on Fig. 1b and Fig. 2b, we conclude that a height difference
of $\Delta d\simeq0.45\,\text{nm}$ is accompanied by a potential
difference of $\Delta V\simeq0.4\,\text{meV}$ (barrier of the condensate
confinement). The latter corresponds to a mirror separation of $D_{0}=-(mc^{2}/n_{0}^{2})\,\Delta d/\Delta V\simeq2.3\,\mu\text{m}$,
which is similar to earlier experiments in the photon BEC system \citep{Kla10b,Kla10}.\vspace{-3mm}

\subsection*{2. Optical medium\vspace{-3mm}
}

The optical medium in our experiment is a water-based solution of
rhodamine 6G (concentration $1\,\text{mMol/l}$) and the thermo-responsive
polymer pNIPAM ($4\,\text{\%}$ mass fraction). To avoid self-quenching
of the dye molecules, we add a small amount of lauryldimethylamine
N-oxide. Due to the addition of pNIPAM, the index of refraction of
our optical medium is expected to be slightly higher than that of
pure water.

A necessary condition for the occurrence of photon Bose-Einstein condensation
in an optical medium is the Kennard-Stepanov (KS) law. The KS law
describes the temperature dependence of absorption $B_{12}(\omega)$
and emission $B_{21}(\omega)$ coefficient of a broadband fluorescent
medium such as a dye solution:
\begin{equation}
\frac{B_{12}(\omega)}{B_{21}(\omega)}=\text{e}^{\frac{\hbar(\omega-\omega_{\text{zpl}})}{kT}}\;\text{.}
\end{equation}
Here, $\omega_{\text{zpl}}$ denotes the zero-phonon line and $T$
describes the temperature of the medium (all experiments were carried
out at room temperature). The KS law goes back to a sub-picosecond
thermalisation process of the vibrational and rotational states of
the dye by collisions with solvent molecules and is well fulfilled
for the dye species in our experiment. It can be shown that multiple
absorption-emission cycles drive a photon gas towards thermal equilibrium
with the optical medium {[}see Klaers et al., \textit{PRL} \textbf{108},
160403 (2012){]}. The KS law is fulfilled, if the emission coefficient
is constant, i.e. $B_{21}(\omega)=\hat{B}$, and the absorption coefficient
increases exponentially with the photon frequency, i.e. $B_{12}=\hat{B}\exp(\hbar(\omega-\omega_{\text{zpl}})/kT)$.
This condition is used in our theoretical modelling and approximately
corresponds to the situation that is realised in experiments.

For optically exciting the dye molecules in the microcavity we use
a pulsed optical parametrical oscillator (OPO) emitting at a wavelength
of $480\,\text{nm}$ with a pulse duration of $5\,\text{ns}$. The
heating of the thermo-responsive polymer is achieved with a $2\,\text{ms}$
laser pulse at a wavelength of $532\,\text{nm}$. The heating pulse
is irradiated onto the sample $50\,\text{ms}$ before the optical
pump pulse. The heating energy is absorbed in an amorphous silicon
layer located below the dielectric stack of one of our mirrors. This
leads to an increase of the local temperature of the optical medium
by a few Kelvin such that it reaches the lower critical solution temperature
(LCST) of pNIPAM in water. The water molecules attached to the polymer
chains are set free and the polymer chains collapse. This initiates
a mass transport that leads to a significant change of the index of
refraction, which, with eq. (\ref{eq:energy_momentum}), translates
into a tunable potential for the photons in the microcavity \citep{Dun17}.

\subsection*{\vspace{-16mm}
}

\section*{APPENDIX B: THEORETICAL METHODS\vspace{-3mm}
}

\subsection*{1. Josephson equations for coupled photon Bose-Einstein condensates\vspace{-3mm}
}

\begin{figure}
\begin{centering}
\ \ \ \ \ \includegraphics[width=8.5cm]{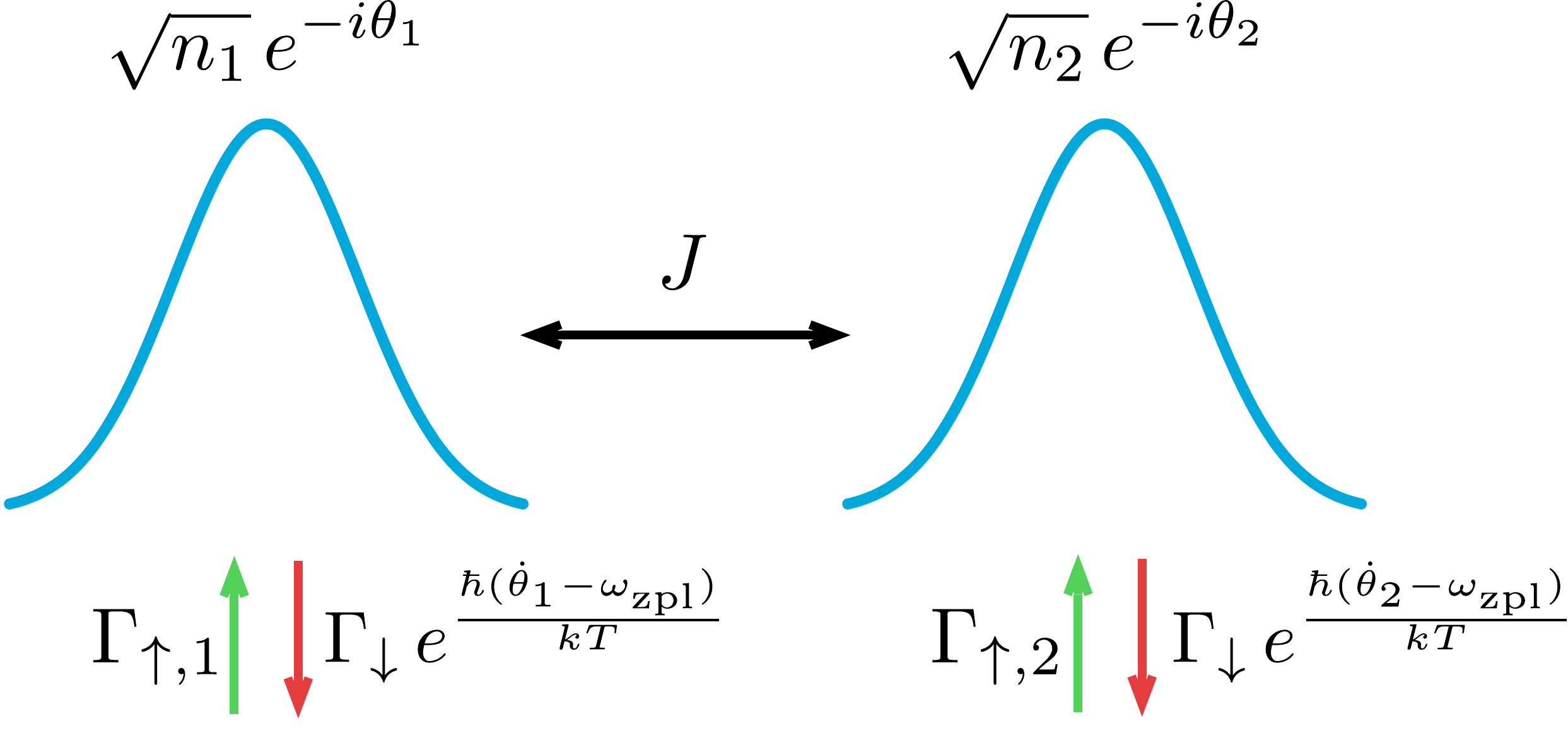}\vspace{-5mm}
\par\end{centering}
\begin{onehalfspace}
\centering{}%
\begin{minipage}[t]{0.95\columnwidth}%
\begin{singlespace}
\begin{center}
\caption{\label{fig:two-coupled-condensates}Two coupled photon Bose-Einstein
condensates with frequency-dependent gain-loss scheme. The latter
follows the Kennard-Stepanov law for fluorescent media and is responsible
for a thermalisation process between the coupled condensate system
and its environment.}
\par\end{center}
\end{singlespace}
\end{minipage}\vspace{-10mm}
\end{onehalfspace}
\end{figure}

The system of two coupled condensates with gain and frequency-dependent
loss is sketched in Fig. \ref{fig:two-coupled-condensates}. It can
be described by two stochastic and dissipative Schr\"odinger equations
\begin{equation}
i\hbar\dot{\psi_{1}}=\hbar\omega_{c,1}\psi_{1}+\frac{i\hbar}{2}\left(\Gamma_{\uparrow,1}-\Gamma_{\downarrow}\exp\left(\frac{\hbar(\dot{\theta}_{1}-\omega_{\text{zpl}})}{kT}\right)\right)\psi_{1}+\hbar J\psi_{2}+\hbar\eta_{1}\psi_{1}\;\;\label{eq:Schroed1}
\end{equation}
\vspace{-7mm}
\begin{equation}
i\hbar\dot{\psi_{2}}=\hbar\omega_{c,2}\psi_{2}+\frac{i\hbar}{2}\left(\Gamma_{\uparrow,2}-\Gamma_{\downarrow}\exp\left(\frac{\hbar(\dot{\theta}_{2}-\omega_{\text{zpl}})}{kT}\right)\right)\psi_{2}+\hbar J\psi_{1}+\hbar\eta_{2}\psi_{2}\;\text{.}\label{eq:Schroed2}
\end{equation}
Here, $\omega_{c,1}$ and $\omega_{c,2}$ describe the bare condensate
frequencies. The parameters $\Gamma_{\uparrow,1}$, $\Gamma_{\uparrow,2}$
and $\Gamma_{\downarrow}$ describe gain and loss rates, while $J$
denotes the coupling of the two condensates. In general, $J$ can
be a complex number. The equations include noise terms $\hbar\eta_{1,2}\psi_{1,2}$
with noise functions $\eta_{1,2}=\eta_{1,2}(t)$. This renders the
time evolution of the system stochastic. The functions $\theta_{1,2}$
denote the phases in the wavefunctions\vspace{-2.5mm}
\begin{equation}
\psi_{1,2}=\sqrt{n_{1,2}}\exp(-i\theta_{1,2})\;\text{,}\label{eq:coherent_state-1}
\end{equation}
We now define the centre frequency of the condensates, namely $\bar{\omega}_{c}=(\omega_{c,1}+\omega_{c,2})/2$,
and rewrite the absorption term in eqs. (\ref{eq:Schroed1}) and (\ref{eq:Schroed2})
as
\begin{align}
\Gamma_{\downarrow}\exp\left(\frac{\hbar(\dot{\theta}_{1,2}-\omega_{\text{zpl}})}{kT}\right) & =\tilde{\Gamma}_{\downarrow}\exp\left(\frac{\hbar(\dot{\theta}_{1,2}-\bar{\omega}_{c})}{kT}\right)
\end{align}
with $\tilde{\Gamma}_{\downarrow}=\Gamma_{\downarrow}\exp(\hbar(\bar{\omega}_{c}-\omega_{\text{zpl}})/kT)$.
Assuming that all occurring frequencies remain close to $\bar{\omega}_{c}$,
i.e. $\hbar|\dot{\theta}_{1,2}-\bar{\omega}_{c}|\ll kT$, we can linearise
the exponential function in the gain-loss term. This leads to
\begin{equation}
i\hbar\dot{\psi_{1}}=\hbar\omega_{c,1}\psi_{1}+\frac{i\hbar}{2}\left(\tilde{\Gamma}_{\uparrow,1}-\tilde{\Gamma}_{\downarrow}\frac{\hbar(\dot{\theta}_{1}-\bar{\omega}_{c})}{kT}\right)\psi_{1}+\hbar J\psi_{2}+\hbar\eta_{1}\psi_{1}\;\ \label{eq:schroed1}
\end{equation}
\vspace{-7mm}
\begin{equation}
i\hbar\dot{\psi_{2}}=\hbar\omega_{c,2}\psi_{2}+\frac{i\hbar}{2}\left(\tilde{\Gamma}_{\uparrow,2}-\tilde{\Gamma}_{\downarrow}\frac{\hbar(\dot{\theta}_{2}-\bar{\omega}_{c})}{kT}\right)\psi_{2}+\hbar J\psi_{1}+\hbar\eta_{2}\psi_{2}\;\text{.}\label{eq:schroed2}
\end{equation}
In the latter, we have introduced the renormalised gain coefficients
$\tilde{\Gamma}_{\uparrow,1}=\Gamma_{\uparrow,1}-\tilde{\Gamma}_{\downarrow}$
and $\tilde{\Gamma}_{\uparrow,2}=\Gamma_{\uparrow,2}-\tilde{\Gamma}_{\downarrow}$.
Using the temporal derivative of eq. (\ref{eq:coherent_state-1}),
we replace the phase velocities via
\begin{equation}
\dot{\theta}_{1,2}=i\left(\frac{\dot{\psi}_{1,2}}{\psi_{1,2}}-\frac{\dot{n}_{1,2}}{2n_{1,2}}\right)\;\text{.}
\end{equation}
Again using eq. (\ref{eq:coherent_state-1}) and separating into real
and imaginary parts (Madelung transformation), we arrive at the Josephson
equations for the coupled photon BEC system. For the evolution of
the condensate phases we find
\begin{equation}
\dot{\theta}_{1}=\omega_{c,1}\text{Re}(\mathcal{A})-\bar{\omega}_{c}[\text{Re}(\mathcal{A})\hspace{-0.5mm}-\hspace{-0.5mm}1]+\frac{1}{2}\text{Im}(\mathcal{A})\left(\frac{\dot{n}_{1}}{n_{1}}\hspace{-0.5mm}-\hspace{-0.5mm}\tilde{\Gamma}_{\uparrow,1}\right)+|\mathcal{A}||J|\sqrt{\frac{n_{2}}{n_{1}}}\,\cos(\theta_{J}\hspace{-0.5mm}+\hspace{-0.5mm}\theta_{\mathcal{A}}\hspace{-0.5mm}+\hspace{-0.5mm}\delta)+\text{Re}(\mathcal{A}\eta_{1})\label{eq:phase1}
\end{equation}
\vspace{-6mm}
\begin{equation}
\dot{\theta}_{2}=\omega_{c,2}\text{Re}(\mathcal{A})-\bar{\omega}_{c}[\text{Re}(\mathcal{A})\hspace{-0.5mm}-\hspace{-0.5mm}1]+\frac{1}{2}\text{Im}(\mathcal{A})\left(\frac{\dot{n}_{2}}{n_{2}}\hspace{-0.5mm}-\hspace{-0.5mm}\tilde{\Gamma}_{\uparrow,2}\right)+|\mathcal{A}||J|\sqrt{\frac{n_{1}}{n_{2}}}\,\cos(\theta_{J}\hspace{-0.5mm}+\hspace{-0.5mm}\theta_{\mathcal{A}}\hspace{-0.5mm}-\hspace{-0.5mm}\delta)+\text{Re}(\mathcal{A}\eta_{2})\label{eq:phase2}
\end{equation}
For the evolution of the particle numbers we derive
\begin{equation}
\dot{n}_{1}=\left[\tilde{\Gamma}_{\uparrow,1}+2\frac{\text{Im}(\mathcal{A})}{\text{Re}(\mathcal{A})}(\omega_{c,1}\hspace{-0.5mm}-\hspace{-0.5mm}\bar{\omega}_{c})\right]n_{1}+\frac{2\,|\mathcal{A}||J|}{\text{Re}(\mathcal{A})}\sqrt{n_{1}n_{2}}\sin(\theta_{J}\hspace{-0.5mm}+\hspace{-0.5mm}\theta_{\mathcal{A}}\hspace{-0.5mm}+\hspace{-0.5mm}\delta)+2\text{Im}(\mathcal{A}\eta_{1})n_{1}\;\:\label{eq:n1}
\end{equation}
\vspace{-8mm}
\begin{equation}
\dot{n}_{2}=\left[\tilde{\Gamma}_{\uparrow,2}+2\frac{\text{Im}(\mathcal{A})}{\text{Re}(\mathcal{A})}(\omega_{c,2}\hspace{-0.5mm}-\hspace{-0.5mm}\bar{\omega}_{c})\right]n_{2}+\frac{2\,|\mathcal{A}||J|}{\text{Re}(\mathcal{A})}\sqrt{n_{1}n_{2}}\sin(\theta_{J}\hspace{-0.5mm}+\hspace{-0.5mm}\theta_{\mathcal{A}}\hspace{-0.5mm}-\hspace{-0.5mm}\delta)+2\text{Im}(\mathcal{A}\eta_{2})n_{2}\;\text{.}\label{eq:n2}
\end{equation}

In these equations, we have introduced a dimensionless complex dissipation
parameter
\begin{equation}
\mathcal{A}=\frac{kT}{kT+\frac{i\hbar}{2}\Gamma_{\downarrow}}=|\mathcal{A}|\text{e}^{i\theta_{\mathcal{A}}}\;\text{.}\label{eq:Z}
\end{equation}
We have furthermore used $\delta=\theta_{1}-\theta_{2}$ and $J=|J|\text{e}^{i\theta_{J}}$.
Omitting the noise terms, we find the equation of motion for the total
particle number $n=n_{1}+n_{2}$:
\begin{equation}
\dot{n}=(\tilde{\Gamma}_{\uparrow,1}n_{1}\hspace{-0.5mm}+\hspace{-0.5mm}\tilde{\Gamma}_{\uparrow,2}n_{2})+2\frac{\text{Im}(\mathcal{A})}{\text{Re}(\mathcal{A})}(\omega_{c,1}n_{1}\hspace{-0.5mm}+\omega_{c,2}n_{2}\hspace{-0.5mm}-\hspace{-0.5mm}\bar{\omega}_{c}n)+\frac{4|\mathcal{A}||J|}{\text{Re}(\mathcal{A})}\sin(\theta_{J}\hspace{-0.5mm}+\hspace{-0.5mm}\theta_{\mathcal{A}})\sqrt{n_{1}n_{2}}\,\cos(\delta)\label{eq:total_particle_number}
\end{equation}

We can draw the following conclusion from this equation:
\begin{enumerate}
\item Equation (\ref{eq:total_particle_number}) implies that a phase difference
of
\begin{equation}
\delta=\begin{cases}
0 & \sin(\theta_{J}+\theta_{\mathcal{A}})>0\\
\pi & \text{otherwise}
\end{cases}\label{eq:delta}
\end{equation}
maximises the gain in total particle number. This holds independent
of the details of the coupling, condensate frequencies and pumping
rates.
\item For condensates with equal frequencies ($\omega_{c,1}=\omega_{c,2}$)
and a given total particle number $n$, the particle number distribution
that maximises the gain is
\begin{equation}
n_{1}=\frac{n}{2}\left[1+\frac{\tilde{\Gamma}_{\uparrow,1}-\tilde{\Gamma}_{\uparrow,2}}{\sqrt{\left(\frac{4|\mathcal{A}||J|}{\text{Re}(\mathcal{A})}\sin(\theta_{J}\hspace{-0.5mm}+\hspace{-0.5mm}\theta_{\mathcal{A}})\right)^{2}+\left(\tilde{\Gamma}_{\uparrow,1}-\tilde{\Gamma}_{\uparrow,2}\right)^{2}}}\right]\;\;\text{with}\;\;n_{2}=n-n_{1}\;\text{.}\label{eq:max_gain}
\end{equation}
For large coupling constants $|J|$ or equal gain ($\tilde{\Gamma}_{\uparrow,1}=\tilde{\Gamma}_{\uparrow,2}$),
the solution reduces to equal population. In other words, the state
$n_{1}=n_{2}=n/2$ with $\delta=0,\hspace{-0.5mm}\pi$ maximises the
gain. This solution is furthermore a dynamical fixed point of the
Josephson equations given in eqs. (\ref{eq:phase1})-(\ref{eq:n2}).
\end{enumerate}

\subsection*{\vspace{-3mm}
2. Potential step model\vspace{-0mm}
}

\begin{figure}
\begin{centering}
~~~~~~~~~~~~~~\includegraphics[width=7cm]{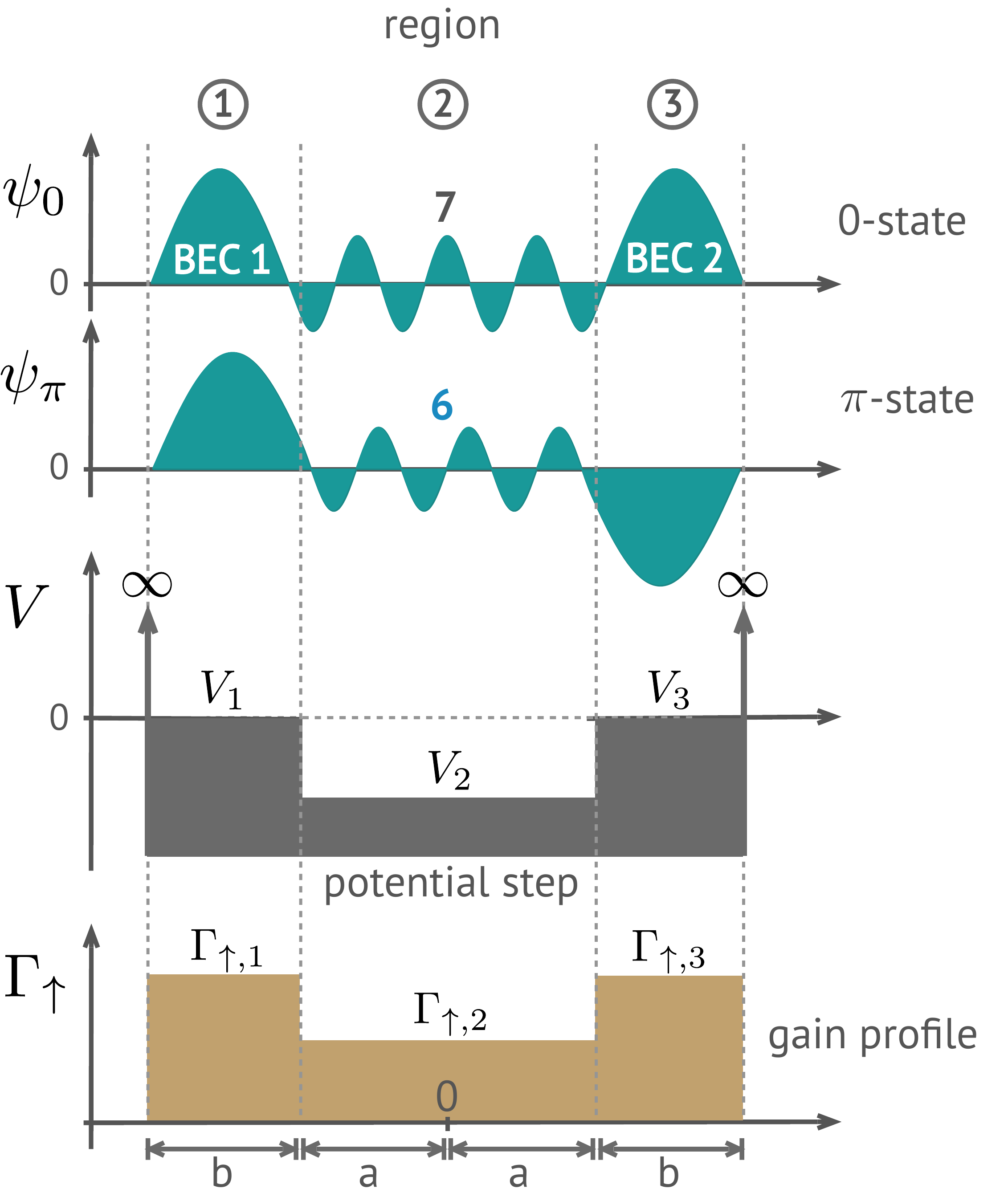}
\par\end{centering}
\begin{centering}
\vspace{-2mm}
\begin{minipage}[t]{0.95\columnwidth}%
\begin{singlespace}
\caption{\label{fig:potential-step-model}Potential step model. The model assumes
that two Bose-Einstein condensates exchange particles via an attractive
potential step. Depending on the potential energy $V$ and gain $\Gamma_{\uparrow}$
the system selects either $0$- or $\pi$-state. We assume spatially
symmetric and piecewise constant functions $V(x)$, $\Gamma_{\uparrow}(x)$.}
\end{singlespace}
\end{minipage}
\par\end{centering}
\begin{onehalfspace}
\vspace{-2mm}
\end{onehalfspace}
\end{figure}

In this section, we analyse a simplifying model, in which two photon
Bose-Einstein condensates exchange particles via an attractive potential
step, see Fig. \ref{fig:potential-step-model}. Using this model,
we will gain insight into the physical mechanisms that drive the state
selection process between symmetric and antisymmetric wavefunctions
in the photonic Josephson junction.

The starting point is a one-dimensional dissipative Schr\"odinger
equation. We use this equation to find symmetric and antisymmetric
wavefunctions that can be associated to the $0$- and $\pi$-state
of a coupled condensate system as discussed in section B.1. The Schr\"odinger
equation now includes terms related to the kinetic and potential energy:\vspace{0mm}
\begin{equation}
i\hbar\frac{\partial\psi}{\partial t}=\hbar\omega_{c}\psi-\frac{\hbar^{2}}{2m}\frac{\partial^{2}\psi}{\partial x^{2}}+V\psi+\frac{i\hbar}{2}\left(\Gamma_{\uparrow}-\Gamma_{\downarrow}\exp\left(\frac{\hbar(\dot{\theta}-\omega_{\text{zpl}})}{kT}\right)\right)\psi\;\text{.}\label{eq:Schroed_stepmodel}
\end{equation}
\vspace{0mm}
We follow a procedure similar as in section B.1. First, we rewrite
the absorption term as
\begin{align}
\Gamma_{\downarrow}\exp\left(\frac{\hbar(\dot{\theta}-\omega_{\text{zpl}})}{kT}\right) & =\tilde{\Gamma}_{\downarrow}\exp\left(\frac{\hbar(\dot{\theta}-\omega_{c})}{kT}\right)
\end{align}
with $\tilde{\Gamma}_{\downarrow}=\Gamma_{\downarrow}\exp(\hbar(\omega_{c}-\omega_{\text{zpl}})/kT)$.
We assume that all occurring energies stay close to the bare condensate
frequency $\omega_{c}$ in the sense that $\hbar(\dot{\theta}-\omega_{c})\ll kT$.
Under this assumption, we linearise the exponential function in the
gain-loss term
\begin{equation}
i\hbar\frac{\partial\psi}{\partial t}=\hbar\omega_{c}\psi-\frac{\hbar^{2}}{2m}\frac{\partial^{2}\psi}{\partial x^{2}}+V\psi+\frac{i\hbar}{2}\left(\tilde{\Gamma}_{\uparrow}-\tilde{\Gamma}_{\downarrow}\frac{\hbar(\dot{\theta}-\omega_{c})}{kT}\right)\psi\;\text{.}
\end{equation}
\vspace{0mm}
In the latter, we have introduced the renormalised gain coefficient
$\tilde{\Gamma}_{\uparrow}=\Gamma_{\uparrow}-\tilde{\Gamma}_{\downarrow}$.
Assuming slowly varying amplitudes in the wavefunction $\psi=\sqrt{n}\exp(-i\theta)$,
we approximate the phase velocity as $\dot{\theta}\simeq i\dot{\psi}/\psi$.
This procedure leads to \vspace{0mm}
\begin{equation}
i\hbar\frac{\partial\psi}{\partial t}=\hbar\omega_{c}\psi-\frac{\hbar^{2}\mathcal{A}}{2m}\frac{\partial^{2}\psi}{\partial x^{2}}+\mathcal{A}V\psi+\frac{i\hbar}{2}\mathcal{A}\tilde{\Gamma}_{\uparrow}\psi\;\text{,}
\end{equation}
\vspace{0mm}
in which we have used the complex dissipation parameter $\mathcal{A}$
as defined in eq. (\ref{eq:Z}). Using $\psi(x,t)=\phi(x)\,\chi(t)$
yields the time-independent Schr\"odinger equation
\begin{equation}
-\frac{\hbar^{2}\mathcal{A}}{2m}\frac{\partial^{2}\phi}{\partial x^{2}}+\mathcal{A}V\phi+\frac{i\hbar}{2}\mathcal{A}\tilde{\Gamma}_{\uparrow}\phi=(E-\hbar\omega_{c})\phi\;\text{.}
\end{equation}
The potential and gain profiles are considered symmetric in space,
i.e. $V(x)=V(-x)$ and $\tilde{\Gamma}_{\uparrow}(x)=\tilde{\Gamma}_{\uparrow}(-x)$.
Furthermore, they are set constant in each of the three regions shown
in Fig. \ref{fig:potential-step-model}. For the symmetric state,
we use the following ansatz
\begin{equation}
\phi_{0}(x)=\begin{cases}
\sin(k_{1}[x\hspace{-0.5mm}+\hspace{-0.5mm}a\hspace{-0.5mm}+\hspace{-0.5mm}b]) & -a\hspace{-0.5mm}-\hspace{-0.5mm}b\leq x\leq-a\\
A\cos(k_{2}x) & \hspace{5mm}-a\leq x\leq a\\
-\sin(k_{1}[x\hspace{-0.5mm}-\hspace{-0.5mm}a\hspace{-0.5mm}-\hspace{-0.5mm}b]) & \hspace{8mm}a\leq x\leq a\hspace{-0.5mm}+\hspace{-0.5mm}b\;\text{.}
\end{cases}\label{eq:wave}
\end{equation}
For the antisymmetric state, we choose
\begin{equation}
\phi_{\pi}(x)=\begin{cases}
\sin(k_{1}[x\hspace{-0.5mm}+\hspace{-0.5mm}a\hspace{-0.5mm}+\hspace{-0.5mm}b]) & -a\hspace{-0.5mm}-\hspace{-0.5mm}b\leq x\leq-a\\
A\sin(k_{2}x) & \hspace{5mm}-a\leq x\leq a\\
\sin(k_{1}[x\hspace{-0.5mm}-\hspace{-0.5mm}a\hspace{-0.5mm}-\hspace{-0.5mm}b])\hspace{2.5mm} & \hspace{8mm}a\leq x\leq a\hspace{-0.5mm}+\hspace{-0.5mm}b\;\text{.}
\end{cases}\label{eq:wave-1}
\end{equation}
Here, the complex wavenumbers $k_{i}$ must follow
\begin{equation}
\ensuremath{k_{i}=\hbar^{-1}\sqrt{2m\,\left(\frac{E-\hbar\omega_{c}}{\mathcal{A}}-V_{i}-\frac{i\hbar}{2}\tilde{\Gamma}_{\uparrow,i}\right)}\;\text{.}}\label{eq:kvec}
\end{equation}
Continuity conditions for $\phi$ and $\partial\phi/\partial x$ at
$x=\pm a$ determine the coefficient $A$. We find
\begin{equation}
A=\begin{cases}
\hspace{3mm}\frac{\sin(k_{1}b)}{\cos(k_{2}a)} & (0\text{-state})\\
-\frac{\sin(k_{1}b)}{\sin(k_{2}a)} & (\pi\text{-state})\;\text{.}
\end{cases}
\end{equation}
Furthermore, the continuity conditions deliver an additional relation
between the wavenumbers. For the two states, these conditions are
given by
\begin{equation}
k_{1}\cot(k_{1}b)=\begin{cases}
\hspace{2.5mm}k_{2}\tan(k_{2}a) & (0\text{-state})\\
-k_{2}\cot(k_{2}a) & (\pi\text{-state})\;\text{.}
\end{cases}
\end{equation}
\begin{figure}
\begin{centering}
\vspace{0mm}
\includegraphics[width=17.5cm]{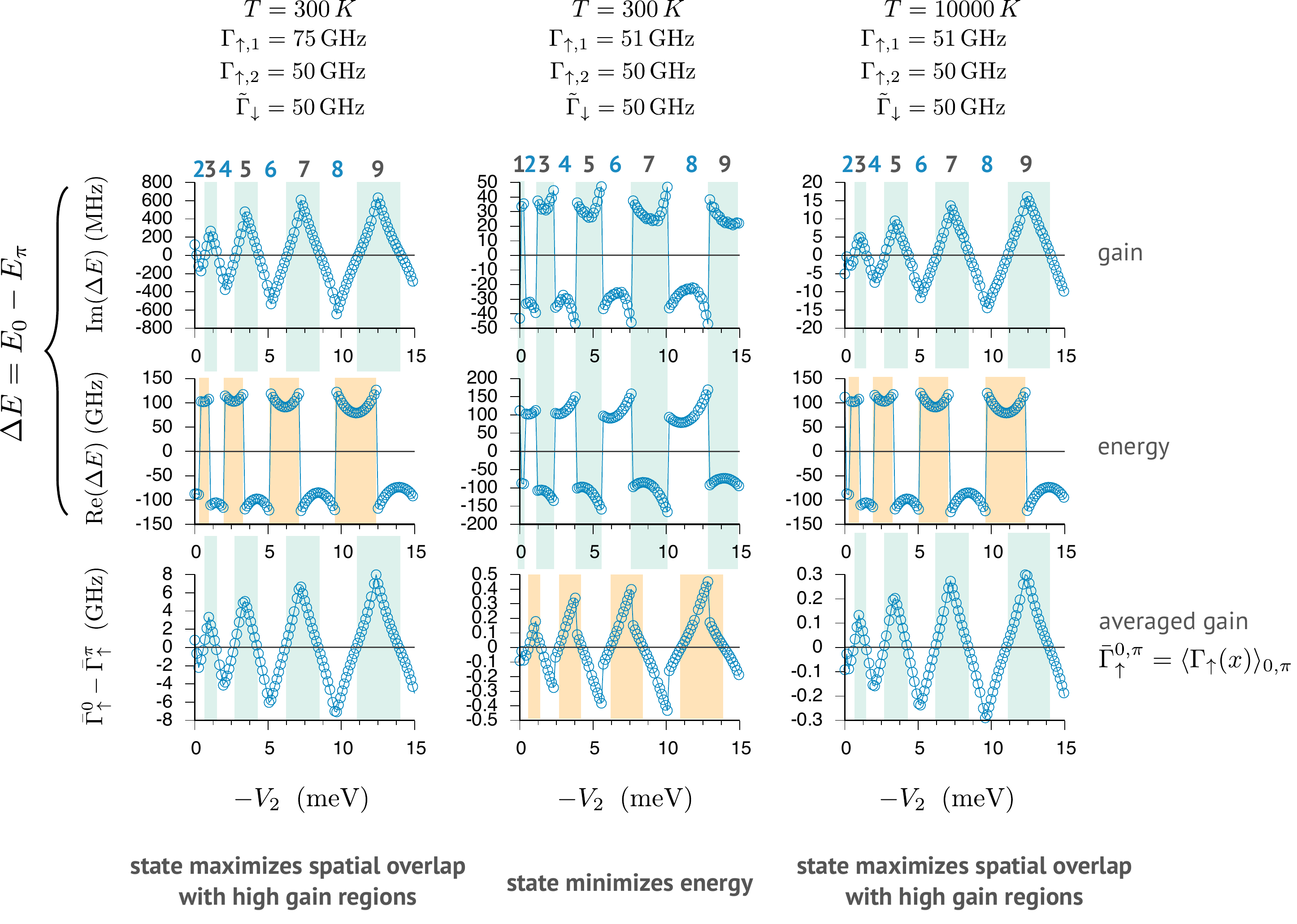}\vspace{0mm}
\par\end{centering}
\begin{centering}
\begin{minipage}[t]{0.95\columnwidth}%
\begin{singlespace}
\caption{\label{fig:potential_step_results}Complex energy difference $\Delta E=E_{0}-E_{\pi}$
as a function of potential step depth $V_{2}$ for three parameter
regimes. For positive $\text{Im}(\Delta E)$, the system predominantly
realises the $0$-state. For negative $\text{Im}(\Delta E)$, the
system chooses to be in the $\pi$-state. For all parameter sets,
the gain periodically switches sign as the depth of the potential
step is increased. For large gain gradients $\Gamma_{\uparrow,1}-\Gamma_{\uparrow,2}$
(left column) or high temperatures $T$ (right column), the system
maximises the spatial overlap with the high gain regions, as indicated
by the quantity $\bar{\Gamma}_{\uparrow}^{0}-\bar{\Gamma}_{\uparrow}^{\pi}$
(see text). For small gradients or low temperatures (middle column),
the system minimizes (the real part of) the energy. Further parameters:
$m=6.5\times10^{-36}\,\text{kg}$, $a=9\,\mu\text{m}$, $b=7\,\mu\text{m}$.}
\end{singlespace}
\end{minipage}
\par\end{centering}
\begin{onehalfspace}
\vspace{0mm}
\end{onehalfspace}
\end{figure}
Together with eq. (\ref{eq:kvec}), these equation determine the allowed
(complex) energies in the system. Since analytical solutions are not
possible, the energy spectrum has to be found numerically. In particular,
we are interested in the energies that belong to the kinetic ground
state of the condensates, which will be called $E_{0}$ and $E_{\pi}$.
These two energies are related to the coupling constant $J$, as defined
in section B.1, in the following way:
\begin{align}
E_{0}-E_{\pi} & =i\hbar\left(\frac{\dot{\psi}_{0}}{\psi_{0}}-\frac{\dot{\psi}_{\pi}}{\psi_{\pi}}\right)\\
 & =\hbar(\dot{\theta}_{0}-\dot{\theta}_{\pi})+\frac{i\hbar}{2}\left(\frac{\dot{n}_{0}}{n_{0}}-\frac{\dot{n}_{\pi}}{n_{\pi}}\right)\\
 & =2\hbar\,\text{Re}(\mathcal{A}J)+i2\hbar\,\text{Im}(\mathcal{A}J)/\text{Re}(\mathcal{A})\;\text{.}
\end{align}
Here, we have used eqs. (\ref{eq:phase1}) and (\ref{eq:total_particle_number})
with $\tilde{\Gamma}_{\uparrow,1}=\tilde{\Gamma}_{\uparrow,2}$, $\omega_{c,1}=\omega_{c,2}$
and consequently $n_{1}=n_{2}$. The solution of this equation in
terms of the coupling constant is
\begin{align}
J & =\frac{\text{Re}(E_{0}-E_{\pi})+i\,\text{Re}(\mathcal{A})\,\text{Im}(E_{0}-E_{\pi})}{2\hbar\mathcal{A}}\;\text{.}
\end{align}
Figure \ref{fig:potential_step_results} shows the complex energy
difference $\Delta E=E_{0}-E_{\pi}$ as a function of the potential
step depth $V_{2}$ for three parameter sets, which represent limiting
cases for the system. For positive imaginary components, i.e. $\text{Im}(\Delta E)>0$,
the $0$-state system has a larger gain than the $\pi$-state. Consequently,
the system will predominantly realise the $0$-state. For negative
$\text{Im}(\Delta E)$, the system chooses the $\pi$-state. For all
parameter sets, the gain periodically switches sign as the depth of
the potential step is increased.

For large gain gradients $\Gamma_{\uparrow,1}-\Gamma_{\uparrow,2}$
(left column) or high temperatures $T$ (right column), the system
primarily maximises the spatial overlap with the high gain region.
The latter is reflected by the fact that the gain curve $\text{Im}(\Delta E)$
correlates with the quantity $\bar{\Gamma}_{\uparrow}^{0}-\bar{\Gamma}_{\uparrow}^{\pi}$,
as shown in the third row of Fig. \ref{fig:potential_step_results}.
Here, $\bar{\Gamma}_{\uparrow}^{0,\pi}=\int_{-a-b}^{-+a+b}\Gamma_{\uparrow}\,|\psi_{0,\pi}|^{2}dx$
(with normalised wavefunctions $\psi_{0,\pi}$) is the spatially averaged
gain. This quantity is positive, if the $0$-state has a larger overlap
with the high gain regions than the $\pi$-state (and is negative
otherwise). For small gradients $\Gamma_{\uparrow,1}-\Gamma_{\uparrow,2}$
or low temperatures $T$ (middle column), the system primarily minimises
the energy, which is reflected by the fact that the gain curve $\text{Im}(\Delta E)$
now correlates with the energy curve $\text{Re}(\Delta E)$. This
behaviour can be understood as a consequence of the thermalisation
process induced by the Kennard-Stepanov gain-loss scheme.\vspace{-3mm}

\subsection*{3. N coupled condensates\vspace{-3mm}
}

Equations (\ref{eq:phase1}), (\ref{eq:phase2}) and (\ref{eq:total_particle_number})
can readily be generalised to $N$ coupled condensates with coupling
constants $J_{ij}$. In the following, we set all condensate frequencies
and pumping rates equal: $\omega_{c,i}=\omega_{c}$ and $\Gamma_{\uparrow,i}=\Gamma_{\uparrow}$.
For the phase velocity, we find:
\begin{equation}
\dot{\theta}_{i}=\omega_{c}+\frac{1}{2}\text{Im}(\mathcal{A})\left(\frac{\dot{n}_{i}}{n_{i}}-\tilde{\Gamma}_{\uparrow}\right)+\sum_{j}|\mathcal{A}||J_{ij}|\sqrt{n_{j}/n_{i}}\,\cos(\theta_{J_{ij}}\hspace{-0.5mm}+\hspace{-0.5mm}\theta_{\mathcal{A}}\hspace{-0.5mm}+\theta_{i}-\hspace{-0.5mm}\theta_{j})\label{eq:phase1-1}
\end{equation}
For the particle number evolution, we obtain
\begin{align}
\dot{n}_{i} & =\tilde{\Gamma}_{\uparrow}n_{i}+\sum_{j}\frac{2|\mathcal{A}||J_{ij}|}{\text{Re}(\mathcal{A})}\sqrt{n_{i}n_{j}}\sin(\theta_{J_{ij}}\hspace{-0.5mm}+\hspace{-0.5mm}\theta_{\mathcal{A}}\hspace{-0.5mm}+\hspace{-0.5mm}\theta_{i}\hspace{-0.5mm}-\hspace{-0.5mm}\theta_{j})\;\text{.}
\end{align}
The total particle number gain $\dot{n}=\sum_{i}\dot{n}_{i}$ is given
by
\begin{align}
\dot{n} & =\sum_{i}\tilde{\Gamma}_{\uparrow}n_{i}+\sum_{i,j}\frac{2|\mathcal{A}||J_{ij}|}{\text{Re}(\mathcal{A})}\sqrt{n_{i}n_{j}}\sin(\theta_{J_{ij}}\hspace{-0.5mm}+\hspace{-0.5mm}\theta_{\mathcal{A}}\hspace{-0.5mm}+\hspace{-0.5mm}\theta_{i}\hspace{-0.5mm}-\hspace{-0.5mm}\theta_{j})\\
 & =\sum_{i}\tilde{\Gamma}_{\uparrow}n_{i}+\sum_{i>j}\frac{4|\mathcal{A}||J_{ij}|}{\text{Re}(\mathcal{A})}\sqrt{n_{i}n_{j}}\sin(\theta_{J_{ij}}\hspace{-0.5mm}+\hspace{-0.5mm}\theta_{\mathcal{A}})\cos(\theta_{i}\hspace{-0.5mm}-\hspace{-0.5mm}\theta_{j})\;\text{,}
\end{align}
which can be written in the form\vspace{-4mm}
\begin{align}
\dot{n} & =\sum_{i}\tilde{\Gamma}_{\uparrow}n_{i}-\sum_{i>j}J_{ij}^{XY}\cos(\theta_{i}\hspace{-0.5mm}-\hspace{-0.5mm}\theta_{j})\;\text{.}\label{eq:n_total}
\end{align}
Here, $J_{ij}^{XY}=-4|\mathcal{A}||J_{ij}|(\text{Re}\mathcal{A})^{-1}\sqrt{n_{i}n_{j}}\sin(\theta_{J_{ij}}\hspace{-0.5mm}+\hspace{-0.5mm}\theta_{\mathcal{A}})$
describes an effective coupling constant. Assuming that all condensates
have the same number of particles, only the phases of the condensates
remain free. In this case, the second term in eq. (\ref{eq:n_total})
exactly corresponds to the Hamiltonian of a classical XY model. The
phase configuration $\{\theta_{i}\}$ that maximises the gain, minimises
the energy of the simulated XY Hamiltonian. This implies that the
coupled BEC system can be used to sample low energy configurations
of the simulated XY Hamiltonian. In general, the assumed equality
of particle numbers is not automatically guaranteed and may require
a feedback scheme \citep{Kal19} (and references therein). Furthermore,
it is not guaranteed that the system relaxes to a single state as
the configuration that maximises the gain is not necessarily a fix
point of the Josephson equations.\vspace{-3mm}

\subsection*{4. Spin glass simulation with photonic Josephson junctions - Numerical
results \vspace{-3mm}
}

In Fig. 5 we use numerical simulations to investigate the dynamics
of the condensation process on a 3-regular planar graph. These simulations
are based on numerical solutions of a two-dimensional, stochastic
and dissipative Schr\"odinger equation:\vspace{-2mm}
\begin{equation}
i\hbar\frac{\partial\psi}{\partial t}=\hbar\omega_{c}\psi-\frac{\hbar^{2}}{2m}\left(\frac{\partial^{2}\psi}{\partial x^{2}}+\frac{\partial^{2}\psi}{\partial y^{2}}\right)+V\psi+\frac{i\hbar}{2}\left(\Gamma_{\uparrow}-\Gamma_{\downarrow}\exp\left(\frac{\hbar\,(\dot{\theta}-\omega_{\text{zpl}})}{kT}\right)-\Gamma_{2}\left|\psi\right|^{2}\right)\psi+\hbar\eta\;\text{.}
\end{equation}
This equation is an extension of eq. (\ref{eq:Schroed_stepmodel})
to the two-dimensional domain. It includes an additional term related
to non-linear losses, namely $-\Gamma_{2}\left|\psi\right|^{2}$,
which allows us to model gain saturation in the system. Moreover,
we added the noise function $\eta=\eta(t)$. We perform the same steps
and approximations as in section B.2 to obtain\vspace{-2mm}
\begin{equation}
i\hbar\frac{\partial\psi}{\partial t}=\hbar\omega_{c}\psi-\frac{\hbar^{2}\mathcal{A}}{2m}\left(\frac{\partial^{2}\psi}{\partial x^{2}}+\frac{\partial^{2}\psi}{\partial y^{2}}\right)+\mathcal{A}V\psi+\frac{i\hbar}{2}\mathcal{A}\left(\tilde{\Gamma}_{\uparrow}-\Gamma_{2}\left|\psi\right|^{2}\right)\psi+\hbar\mathcal{A}\eta\;\text{.}\label{eq:Schroedinger_numerical}
\end{equation}
Numerical solutions of this equation are obtained with the Runge-Kutta
method (4th order) with constant time steps. The open source software
library ViennaCL is used to perform the computations on a fast GPU.
We have verified that our numerical results reproduce known analytical
results in a series of test cases. The ground state problem shown in Fig. 5 is mapped onto the potential
landscape in the microcavity system by introducing a triangular lattice
potential, in which the lattice sites are connected by attractive
step potentials. To simulate antiferromagnetic couplings ($J_{ij}^{XY}=-1$),
the depth of the steps is chosen to favour the formation of $\pi$-states.
The gain profile $\Gamma_{\uparrow}(x,y)$ follows the lattice geometry.
The spatial profile of the loss (absorption) is assumed to be homogeneous
$\Gamma_{\downarrow}(x,y)=1\,\text{THz}$. We furthermore assume $T=300\,\text{K}$,
which sets the dissipation parameter $\mathcal{A}$ given in eq. (\ref{eq:Z}).
Condensate frequency and zero-phonon line are considered equal: $\omega_{c}=\omega_{\text{zpl}}$.
The effective photon mass in the simulation is $m=6.5\cdot10^{-36}\text{\,\text{kg}}$.

~
\end{document}